\documentclass[%
 aip,
 jcp,% groupedaddress,
 amsmath,amssymb,
reprint,%
]{revtex4-1}

\usepackage[title]{appendix}
\usepackage{dsfont}
\usepackage{tikz}
\usetikzlibrary{shapes,arrows}
\usepackage{makecell}
\usepackage{colortbl}
\usepackage{graphicx}% Include figure files
\usepackage{dcolumn}% Align table columns on decimal point
\usepackage{bm}% bold math
\usepackage{natbib}
\setcitestyle{square}
\usepackage{ulem}
\usepackage{setspace}
\usepackage{mathtools}

\DeclarePairedDelimiter\floor{\lfloor}{\rfloor}

\definecolor{basilgreen}{rgb}{0.25, 0.42, 0.11}

\begin{document}

%\preprint{AIP/123-QED}

%\title[multistate ring-polymer instantons]{On the multistate ring-polymer instanton}% Force line breaks with \\

\title[Multistate ring polymer instanton]{Multistate ring polymer instantons and nonadiabatic reaction rates}% Force line breaks with \\
%\thanks{Footnote to title of article.}

\author{Srinath Ranya} \author{Nandini Ananth} \email{ananth@cornell.edu.}
%Authors' institution and/or address%\\This line break forced with \textbackslash\textbackslash
%}%

% \homepage{http://www.Second.institution.edu/~Charlie.Author.}
\affiliation{Department of Chemistry and Chemical Biology, 
Cornell University, Ithaca, New York, 14853, USA}
%
%Second institution and/or address%\\This line break forced% with \\
%}%

\date{\today}% It is always \today, today,

\begin{abstract}

    We present two multistate ring polymer instanton (RPI) formulations, 
        both obtained from an exact path integral representation of 
	the quantum canonical partition function for multistate	systems.
    The two RPIs differ in their treatment of the electronic degrees of freedom; 
    whereas the Mean-Field (MF)-RPI averages over the electronic state contributions, 
        the Mapping Variable (MV)-RPI employs 
    explicit continuous Cartesian variables to represent the electronic states. 
    
    We compute both RPIs for a series of 
    model two-state systems coupled to a single nuclear mode 
    with electronic coupling values chosen to describe dynamics 
    in both adiabatic and nonadiabatic regimes. 
	We show that the MF-RPI for symmetric
	systems are in good agreement with previous literature, and we show that our numerical techniques are
	robust for systems with non-zero driving force. The nuclear MF-RPI and the 
	nuclear MV-RPI are similar, but the MV-RPI uniquely reports on the changes 
	in the electronic state populations along the instanton 
	path. In both cases, we analytically demonstrate the existence of a 
    zero-mode and we numerically find that these 
    solutions are true instantons with a single unstable mode 
    as expected for a first order saddle point. Finally, we use the 
    MF-RPI to accurately calculate rate constants for adiabatic and 
    nonadiabatic model systems with the coupling strength varying 
    over three orders of magnitude.    
\end{abstract}

%Use showkeys class option if keyword
\maketitle

%\begin{quotation}
%The ``lead paragraph'' is encapsulated with the \LaTeX\ 
%\verb+quotation+ environment and is formatted as a single paragraph before the first section heading. 
%(The \verb+quotation+ environment reverts to its usual meaning after the first sectioning command.) 
%Note that numbered references are allowed in the lead paragraph.
%
%The lead paragraph will only be found in an article being prepared for the journal \textit{Chaos}.
%\end{quotation}

\section{\label{sec:level1}Introduction}
Nonadiabatic charge and energy transfer processes are the key step in 
the functioning of many biological and chemical systems.~\cite{Marcus1985,Gray1996,Marcus1993,Cukier1998, Reece2009} As such,
the computation of nonadiabatic reaction rates has been the subject
of a great deal of interest for over two decades, leading to the development of several 
nonadiabatic dynamic approaches. An alternate low-cost strategy to compute
rates for adiabatic and nonadiabatic processes is the computation of 
so-called `instantons'.~\cite{Miller1975, Langer2000} 
An instanton is a periodic orbit in imaginary time on an inverted potential 
energy surface~\cite{Coleman1979} and is, typically, the trajectory that contributes most to
the flux-side correlation function.~\cite{Chapman1975, Miller1983}

Semiclassical instanton rate theory has been employed with a great deal of success 
for the computation of adiabatic reaction rates.~\cite{Callan1977, Coleman1977, Kryvohuz2011, Miller2003} 
More recently, the ring polymer instanton (RPI) method 
based on the path integral formulation of quantum mechanics
has been developed~\cite{Richardson2009} and it's connections with semiclassical
theory established.~\cite{Althorpe2011, Richardson2016a}
Early work towards nonadiabatic rates via an instanton formulation
extended the semiclassical instanton approach to multistate
systems and was successfully used to calculate both adiabatic and nonadiabatic 
rates for symmetric systems.~\cite{Schwieters1998, Schwieters1999, Jang2001} 
Other methods to compute instanton rates 
valid only in the nonadiabatic or weak-coupling 
limit include a saddle point approximation to the flux-flux 
correlation function by Wolynes,~\cite{Wolynes1987} recently
shown to be accurate in the inverted Marcus regime;~\cite{Lawrence2018}  
and a nonadiabatic instanton obtained by extending  
Gutzwiller's work~\cite{Gutzwiller1967,Gutzwiller1971} to imaginary time 
and energy-matching two single surface instantons at the point 
of crossing.~\cite{Richardson2015a, Richardson2015b} 

In this paper, we numerically calculate multistate RPIs
that are first order saddles, we obtain analytic expressions 
for the zero mode, and we derive an expression to calculate rate 
constants applicable to both nonadiabatic and adiabatic processes. 
We derive two expressions for multistate RPIs. 
The first, a Mean-Field 
(MF)-RPI, is obtained by finding the stationary path in imaginary time 
from the exact mean-field path integral representation of the 
quantum canonical partition function. The MF-RPI closely follows 
previously proposed nonadiabatic instantons~\cite{Schwieters1998} \textemdash\,
it is accurate for both nonadiabatic and adiabatic processes but 
does not explicitly report on transitions between electronic states
along the instanton path.
An equivalent representation of the canonical partition function of multistate
systems can be obtained employing continuous Cartesian variables for both 
the nuclear degrees of freedom and the electronic state variables using 
the Meyer-Miller-Stock-Thoss mapping protocol.~\cite{Stock1997, Meyer1979a, Ananth2010} 
Such a mapping-variable (MV) formulation has been shown to be particularly useful
in developing approximate nonadiabatic dynamic methods.~\cite{Huo2011, Kelly2012, Ananth2013, Chowdhury2017, Miller2017}
Here, we compute the continuous mapping-variable (MV)-RPI that explicitly
includes both nuclear positions and electronic state populations
along the instanton path.

We numerically compute the MF-RPI and the MV-RPI for a series 
of model two-state systems coupled to a single nuclear degree 
of freedom using the Limited-memory Broyden-Fletcher-Goldfarb-Shanno 
minimization algorithm with box constraints (L-BFGS-B).~\cite{Zhu1997}
We show that both nuclear instantons
correctly describe the physics of a nonadiabatic transition.
We further demonstrate that the MV-RPI uniquely reports
on changes in electronic state populations in both the 
adiabatic and nonadiabatic regimes without any
assumptions about the nuclear positions at 
which electronic state transitions occur. Finally, 
we derive an MF-RPI rate expression and 
show that the resulting rate constants 
agree well with Fermi's Golden Rule (FGR) rates
for nonadiabatic model systems and with 
single surface RPI rates in the adiabatic regime. 

The paper is organized is as follows: In Sec.~\ref{sec:level2} we provide an 
overview of the MF-RPI and the MV-RPI and introduce the MF-RPI rate
expression. We provide a brief description of the model systems in Sec.~\ref{sec:level3}. Implementation details are provided in Sec.~\ref{sec:level4}, 
and we present our results and conclusions 
in Sec.~\ref{sec:level5} and Sec.~\ref{sec:level6},
respectively.

\section{\label{sec:level2}Theory}

\subsection{\label{ssec:level2A} Single Surface Ring Polymer Instanton }
In general, the Hamiltonian corresponding to a single surface 
system with $f$ nuclear degrees of freedom (dofs) can be written as
\begin{eqnarray}
    \hat{H}(\hat{\mathbf{R}},\hat{\mathbf{P}}) & \equiv & \dfrac{\hat{\mathbf{P}}^{T}\hat{\mathbf{P}}}{2M} + \mathrm{V}(\hat{\mathbf{R}}),
\end{eqnarray} 
where, $\hat{\mathbf{R}}$ and $\hat{\mathbf{P}}$ 
represent the positions and momenta of the nuclear dofs,
$M$ is the nuclear mass, 
and $\mathrm{V}(\hat{\mathbf{R}})$ is the potential. 
The path integral expression for the quantum canonical 
partition function is obtained from the trace of the 
Boltzmann operator by inserting multiple
copies of identity in the form of a complete set of nuclear
position states,
\begin{eqnarray}
\nonumber
\mathcal{Z} & = & \mathrm{Tr}[e^{-\beta\hat{H}}] \\
& \propto & \lim_{N\rightarrow \infty} \int d\{ \mathbf{R}_{\alpha} \} e^{-\beta V_{\mathrm{RP}}}, 
\label{ZSSRP}
\end{eqnarray}
where we omit pre-multiplicative constants, $\beta~=~1/k_BT$
and we use the notation 
$\int d\left\{ \mathbf{R}_\alpha \right\}~\equiv~\int d\mathbf{R}_1\ldots\int d\mathbf{R}_N$.
In Eq.~\ref{ZSSRP}, we define the isomorphic ring polymer potential as 
$V_\mathrm{RP}~=~U(\{\mathbf{R}_\alpha\})~+~(1/N)\sum_\alpha \mathrm{V}(\mathbf{R}_\alpha)$, 
where the inter-bead interaction potential is
\begin{equation}
    U(\left\{ \mathbf{R}_\alpha \right\}) = \dfrac{1}{N}\sum_{\alpha} \dfrac{M}{2\beta_{N}^2}(\mathbf{R}_{\alpha} - \mathbf{R}_{\alpha+1})^T(\mathbf{R}_{\alpha} - \mathbf{R}_{\alpha+1}),
\label{springsEq}
\end{equation}
and $\beta_N=\beta/N$

The RPI is a discretized approximation 
to the instanton path, and is a first order saddle point on the 
ring polymer potential.\cite{Richardson2009} 
It is determined by 
setting the gradient of the isomorphic classical potential $V_\mathrm{RP}$ to zero, 
\begin{eqnarray}
    \frac{\partial V_\mathrm{RP}}{\partial [\mathbf{R}_{\alpha}]_{i}} = 0,  
    \label{SSeqns} 
\end{eqnarray}
where bead index $\alpha = 1\ldots N$ and the nuclear dofs 
are indexed by $i=1\ldots f$. In the absence of an analytic
solution, the RPI is numerically obtained as the 
simultaneous solution of the $fN$ equations in Eq.~\ref{SSeqns}.

We note that the ring polymer potential, $V_\mathrm{RP}$, is invariant to 
cyclic permutation of the beads that define the RPI path
and in the $N\rightarrow\infty$ it reproduces the familiar result \textemdash\, the invariance of the instanton path action to imaginary time translation. 

\subsection{\label{ssec:level2B} Canonical Partition Function for Multistate Systems}

The potential for a multi-state system with 
$\mathcal{K}$ electronic states and $f$ nuclear dofs in
the diabatic representation is,
\begin{eqnarray}
%\hat{H}(\hat{\mathbf{R}},\hat{\mathbf{P}}) & \equiv & \dfrac{\hat{\mathbf{P}}^{T}\hat{\mathbf{P}}}{2M} + \mathbf{V}(\hat{\mathbf{R}}) \\
\mathbf{V}(\hat{\mathbf{R}})  =  \sum_{n,m=1}^{\mathcal{K}} |\psi_{n} \rangle V_{nm}(\hat{\mathbf{R}}) \langle \psi_{m} |,
\end{eqnarray} 
where $\{|\psi_{n}\rangle\}$ are the diabatic electronic states, the diagonal matrix elements
$V_{nn}(\hat{\mathbf{R}})$ are the potentials corresponding to the $n^{\mathrm{th}}$ state, 
and the off-diagonal matrix elements $V_{nm}(\hat{\mathbf{R}})$ 
describe the electronic couplings between states $n$ and $m$. 
The canonical partition function is expressed as a trace over 
the Boltzmann operator,
\begin{eqnarray}
\nonumber
\mathcal{Z} & = & \mathrm{Tr}_{ne}[e^{-\beta\hat{H}}] \\
& \propto & \lim_{N\rightarrow \infty} \int d\{ \mathbf{R}_{\alpha} \} \label{ZNElec1}
e^{-\beta U(\left\{ \mathbf{R}_\alpha \right\})} 
\mathrm{Tr}_{e}\left[\prod_{\alpha=1}^{N} e^{-\beta_{N} \mathbf{V}(\mathbf{R}_{\alpha})}\right]\;
\label{ZNElec2}
\end{eqnarray}
where the subscripts $n$ and $e$ indicate that the trace 
is evaluated over the nuclear and electronic dofs respectively,
and $U(\left\{ \mathbf{R}_\alpha \right\})$ is the inter-bead 
potential previously defined in Eq.~\ref{springsEq}.
We obtain Eq.~\ref{ZNElec2}, by evaluating the trace 
over the nuclear dofs in the 
position basis.
The trace over electronic dofs can be evaluated in a number of ways;~\cite{Ananth2010, Ananth2013, Richardson2013,  Duke2017, Chowdhury2017} here we explore the MF and one 
MV formulation with explicit electronic state variables.

\subsubsection{\label{ssec:level2B1} Mean-Field Representation}
The MF representation of the canonical partition 
function~\cite{Schwieters1999, Schwieters1998, Schmidt2007, Hele2011, Ananth2013, Duke2017}
is obtained by introducing $N$ copies of the identity,
\begin{equation}
    \mathds{1} = \sum_{n} | \psi_{n} \rangle \langle \psi_{n} |, 
\label{resIdMF}  
\end{equation}
in Eq.~\ref{ZNElec2} and evaluating the trace to obtain
~\cite{Pierre2017}
\begin{eqnarray}
    \mathrm{Tr}_{e}\left[\prod_{\alpha=1}^{N} e^{-\beta_{N} \mathbf{V}(\mathbf{R}_{\alpha})}\right] 
    & = & \mathrm{Tr}\left[ \prod_{\alpha=1}^{N} \mathcal{M}(\mathbf{R}_{\alpha},\mathbf{R}_{\alpha+1})\right] 
    \equiv \Gamma_{\mathrm{MF}}, \label{gamma}  \notag \\ 
\end{eqnarray}
where the matrix elements 
$\mathcal{M}(\mathbf{R}_{\alpha},\mathbf{R}_{\alpha+1})$ are
\begin{eqnarray}
    \mathcal{M}_{nn} & = & e^{-\beta_{N}/2[V_{nn}(\mathbf{R}_{\alpha})+V_{nn}(\mathbf{R}_{\alpha})]}, \notag \\
\mathcal{M}_{nm} & = & -\beta_{N}/4  \left[  \makecell{V_{nm}(\mathbf{R}_{\alpha}) + V_{nm}(\mathbf{R}_{\alpha+1}) } \right] \notag \\ & \times & \left[ \makecell{e^{-\beta_{N}/2[V_{nn}(\mathbf{R}_{\alpha})+V_{nn}(\mathbf{R}_{\alpha+1})]} \vspace*{0.2cm}\\ + e^{-\beta_{N}/2[V_{mm}(\mathbf{R}_{\alpha})+V_{mm}(\mathbf{R}_{\alpha+1})]} } \right].
\end{eqnarray} 
The quantum canonical partition function 
in the MF representation is then
\begin{eqnarray}
\mathcal{Z}_\mathrm{MF} & \propto & \lim_{N\rightarrow \infty} \int d\{ \mathbf{R}_{\alpha} \} e^{-\beta V_{\mathrm{MF}}(\{ \mathbf{R}_{\alpha} \})} \mathrm{sgn}(\Gamma_{\mathrm{MF}})
\end{eqnarray}
where $\mathrm{sgn}(\Gamma_{\mathrm{MF}})$ 
ensures that the partition function is positive definite, 
and we have omitted pre-multiplicative constants. 
The effective MF ring polymer potential is 
\begin{eqnarray}
	V_{\mathrm{MF}}(\{ \mathbf{R}_{\alpha} \}) = U(\left\{ \mathbf{R}_\alpha \right\}) - \dfrac{1}{\beta} \ln | \mathrm{Re}(\Gamma_{\mathrm{MF}}) |,
\end{eqnarray}
where $\Gamma_{\mathrm{MF}}$, given in Eq.~\ref{gamma}, averages over the 
electronic state configurations of the ring polymer making this 
a `mean-field' formulation.

\subsubsection{\label{ssec:level2B2} Mapping Variable Representation}
We introduce continuous Cartesian variables for the electronic states 
using the MMST mapping protocol.~\cite{Stock1997, Meyer1979a, Meyer1979b} 
Specifically, the $\mathcal{K}$ diabatic electronic states are mapped to a 
singly excited oscillator (SEO) basis where $\mathcal{K}-1$ 
harmonic oscillators are in the ground state
and one oscillator (the $n^\mathrm{th}$) is 
in the first excited state,
\begin{eqnarray}
| \psi_{n} \rangle \langle \psi_{m} |  & \rightarrow & \hat{a}_{n}^{\dagger}\hat{a}_{m} \notag\\
| \psi_{n} \rangle & \rightarrow & | 0_{1},\dots,1_{n},\dots,0_{\mathcal{K}}\rangle \equiv | n\rangle 
\end{eqnarray}
The resolution of identity in the 
electronic variables ($\mathbf{x}$) is~\cite{Ananth2010}
\begin{equation}
\mathds{1} = \int d\mathbf{x} \,
|\mathbf{x}\rangle \langle \mathbf{x}| \mathcal{P},
\label{IdMV}
\end{equation}
where the projection operator 
$\mathcal{P} = \sum_{n} | n \rangle \langle n |$ constrains the electronic 
coordinates to the SEO subspace. 

Introducing multiple copies of this identity and evaluating the electronic trace in Eq.~{\ref{ZNElec2}}, we obtain an expression for the partition function:~\cite{Ananth2010}
\begin{eqnarray}
    \mathcal{Z}_{\mathrm{MV}}  \propto  \lim_{N\rightarrow \infty} 
    \int d\{ \mathbf{R}_{\alpha} \}  \int d\{ \mathbf{x}_{\alpha} \} 
    e^{-\beta V_{\mathrm{MV}}} \mathrm{sgn}(\Gamma_{\mathrm{MV}}),\;\;\;\;\;
\end{eqnarray}
where 
\begin{equation}
    \Gamma_{\mathrm{MV}}=\mathrm{Tr}\left[ \prod_{\alpha=1}^{N} \mathcal{X}_{\alpha} \mathcal{M}(\mathbf{R}_{\alpha},\mathbf{R}_{\alpha+1}) \right],
\end{equation}
the matrix 
$\mathcal{X}_{\alpha} = \mathbf{x}_{\alpha} \otimes \mathbf{x}_{\alpha}^{T}$,
and the MV ring polymer potential is,
\begin{eqnarray}
    V_{\mathrm{MV}}(\{ \mathbf{R}_{\alpha}\}) =  U(\{ \mathbf{R}_{\alpha}\}) + \dfrac{1}{\beta} \sum_{\alpha}\mathbf{x}_{\alpha}^T\mathbf{x}_{\alpha} - \dfrac{1}{\beta} \ln | \mathrm{Re}(\Gamma_{\mathrm{MV}}) |. \notag \\
\end{eqnarray}
\subsection{\label{ssec:level2C} Multistate Ring Polymer Instanton}
The multistate (MS) RPI is obtained by finding the MF or MV ring polymer configuration that is a first order saddle
point on the corresponding potential energy surface. 
In the MF-RP formulation, the $fN$ equations that must be solved simultaneously are
\begin{equation}
\frac{\partial V_{\mathrm{MF}}}{\partial [\mathbf{R}_{\alpha}]_{i}} = 0  \label{MFeqns}
\end{equation}
where $\alpha$ is the bead number index and index $i$ runs over the nuclear dofs.
The MV-RPI is obtained by solving $(f+\mathcal{K})N$ equations simultaneously
\begin{eqnarray}
\frac{\partial V_{\mathrm{MV}}}{\partial [\mathbf{R}_{\alpha}]_{i}} & = & 0 \;\;\text{and} \label{MVeqns1}  \\
\frac{\partial V_{\mathrm{MV}}}{\partial [\mathbf{x}_{\alpha}]_{j}} & = & 0,  \label{MVeqns2}
\end{eqnarray}
where the index $j$ runs over the electronic states.
\subsection{\label{ssec:level2D} Zero mode of the instanton}
A true instanton solution is a first order saddle with one negative eigenvalue and 
a zero eigenvalue due to the invariance of the action under imaginary time translation.~\cite{Callan1977} 
In the case of the single surface RPI, the zero eigenvalue mode 
corresponds to the invariance of the isomorphic ring polymer potential 
under cyclic permutation of the beads.~\cite{Richardson2009} 
Here, we establish the existence of a negative eigenvalue numerically for both the MF-RPI and MV-RPI, and 
we analytically find the zero mode in each case.
The action for multistate systems, $\mathcal{S}_{\mathrm{MS}}$, is obtained by taking 
the continuum limit of the isomorphic ring polymer potential, $V_\mathrm{MS}$, 
where $\mathrm{MS}=\{\mathrm{MF,MV}\}$,
\begin{eqnarray}
\mathcal{S}_{\mathrm{MS}}  = && \int_{0}^{\beta} d\tau 
   \left[ \dfrac{M}{2}\left( \dfrac{d \mathbf{X}(\tau)}{d\tau}\right)^2 
   + \mathrm{V_{MS}}[\mathbf{X}(\tau)]\right],
\end{eqnarray}
and $\mathbf{X}(\tau)$ is a vector of all the dofs (nuclear only for the MF-RPI, 
and nuclear and electronic for the MV-RPI) in imaginary time, $\tau$.
The MS-instanton solution is obtained by setting 
the first variation of the action to zero, 
\begin{eqnarray}
\delta \mathcal{S}_{\mathrm{MS}} =
-M \dfrac{d^2 \mathbf{X}(\tau)}{d\tau^2} + \nabla_{\mathbf{X}} \mathrm{V_{MS}}[\mathbf{X}(\tau)] = 0 \label{ms-eom} 
\end{eqnarray}
The second variation of the action, $\delta^2 \mathcal{S}_{\mathrm{MS}}$, 
is the stability matrix ($\Lambda_{\mathrm{MS}}$) which incorporates 
the effects of path fluctuations about the MS instanton,
\begin{eqnarray}
\delta^2 \mathcal{S}_{\mathrm{MS}} & = & \int_{0}^{\beta} d\tau\ \delta \mathbf{X}(\tau)^{T}  \Lambda_{\mathrm{MS}}
\delta \mathbf{X}(\tau) \notag \\
\Lambda_{\mathrm{MS}} & = & \left[ - M \dfrac{d^2 }{d\tau^2} + \nabla_{\mathbf{X}}^{T} \nabla_{\mathbf{X}}\mathrm{V_{MS}}  \right]
\end{eqnarray}
Differentiating Eq.~\ref{ms-eom} with respect to
imaginary time,
\begin{eqnarray}
\nonumber
\dfrac{d}{d\tau} \left[ - M \dfrac{d^2 \mathbf{X}(\tau)}{d\tau^2} + \nabla_{\mathbf{X}}\mathrm{V_{MS}}  \right] =  \Lambda_{\mathrm{MS}}\dot{\mathbf{X}}(\tau) = 0 \times \dot{\mathbf{X}}(\tau) \notag \\
\end{eqnarray}
we find the zero-mode of the MS-instanton 
corresponds to a velocity mode in all the system dofs.
Details of the derivation along with the 
stability matrices corresponding to the MF-RPI and MV-RPI are provided 
in Appendices~\ref{sec:AppendA} and~\ref{sec:AppendB}, respectively. 

\subsection{\label{ssec:level2E} Reaction Rate from MF-RPI}
We can express the MF RPI rate as~\cite{Richardson2009, Althorpe2011, Benderskii1994} 
\begin{eqnarray}
k_{\mathrm{MF-RPI}} & \approx & \dfrac{2}{\beta} 
\dfrac{\mathcal{Z}_{b}}{\mathcal{Z}_{r}}  \\
& = & \dfrac{2e^{-\beta_{N}V_{\mathrm{MF}}
(\tilde{\mathbf{R}}_{\alpha})} }{\mathcal{Z}_{r}\beta} 
\left( \dfrac{M}{2\pi\beta_{N}} \right)^{\frac{N}{2}} \notag \\
&\times & \int d\{ \mathbf{R}_{\alpha} \} 
e^{-\beta_{2N}V''_{\mathrm{MF}}(\tilde{\mathbf{R}}_\alpha)
\;(\mathbf{R}_{\alpha}-\tilde{\mathbf{R}}_{\alpha})^2},
\label{eq:mfrpi_rate}
\end{eqnarray}
where $\{ \tilde{\mathbf{R}}_\alpha \}$ represents the MF-RPI configuration,
$\mathcal{Z}_{b}$ is the barrier partition function, $\mathcal{Z}_{r}$ is the reactant partition function, and the second derivative $V''_{\mathrm{MF}}$ is evaluated
at the MF-RPI configuration. 
In Eq.~\ref{eq:mfrpi_rate}, the last line is obtained by 
Taylor expanding the MF-PES about the MF-RPI solution and truncating
to second order.
The diagonalization of the hessian $V''_{\mathrm{MF}}(\{ \tilde{\mathbf{R}}_\alpha \})$ 
yields $N$ eigenvalues ($M\lambda_{\alpha}^2$): 
\begin{eqnarray}
k_{\mathrm{MF-RPI}} & = & \dfrac{2e^{-\beta_{N}V_{\mathrm{MF}}(\tilde{\mathbf{R}}_{\alpha})} }{\mathcal{Z}_{r}\beta} \left( \dfrac{M}{2\pi\beta_{N}} \right)^{\frac{N}{2}} \notag \\
&\times & \int d\{ \mathbf{s}_{\alpha} \} e^{-\beta_{2N}\sum_{\alpha} M\lambda_{\alpha}^2\mathbf{s}_{\alpha}^2}
\label{eq:mfrpi_int}
\end{eqnarray}
The Hessian evaluated at the MF-RPI configuration has 
one negative eigenvalue, $\lambda_{1} < 0 $, and a zero eigenvalue, 
$\lambda_{2} \approx 0$, for a bead-converged calculation. 
We use this to evaluate the integral in Eq.~\ref{eq:mfrpi_int} 
\textemdash\, mode $\mathbf{s}_{1}$ is integrated by analytically 
continuing the Gaussian and performing the integral over 
the positive part of the imaginary axis, mode $\mathbf{s}_{2}$ 
is integrated out analytically as are the remaining $(N-2)$ 
Gaussian integrals. We note that this derivation closely follows
that employed in deriving the single surface RPI rate.~\cite{Richardson2009} 
The resulting expression for the MF-RPI rate constant is then,
\begin{eqnarray}
k_{\mathrm{MF-RPI}} & \approx & \dfrac{e^{-\beta_{N}V_{\mathrm{MF}}
(\tilde{\mathbf{R}}_{\alpha})} }{\mathcal{Z}_{r} \beta_{N}} 
\left(\dfrac{M z_{N}}{2\pi\beta_{N}}\right)^{\frac{1}{2}} 
\sideset{}{'}
\prod_{\alpha=1}^{N}\dfrac{1}{\beta_{N}|\lambda_{\alpha}|} , \;\;\;\;\;\; 
\label{eq:kMF}
\end{eqnarray}
where the $^\prime$ on the product indicates that $\alpha=2$ is excluded, and $z_{N} = \sum_{\alpha} (\tilde{\mathbf{R}}_{\alpha} - \tilde{\mathbf{R}}_{\alpha + 1})^2$. 
This expression follows a previously derived nonadiabatic instanton rate,
with the primary difference being the method 
used to find the instanton.~\cite{Schwieters1998}

\section{\label{sec:level3} Model Systems}
\begin{figure}[htp]
\centering
\includegraphics[scale=0.3,angle=-90]{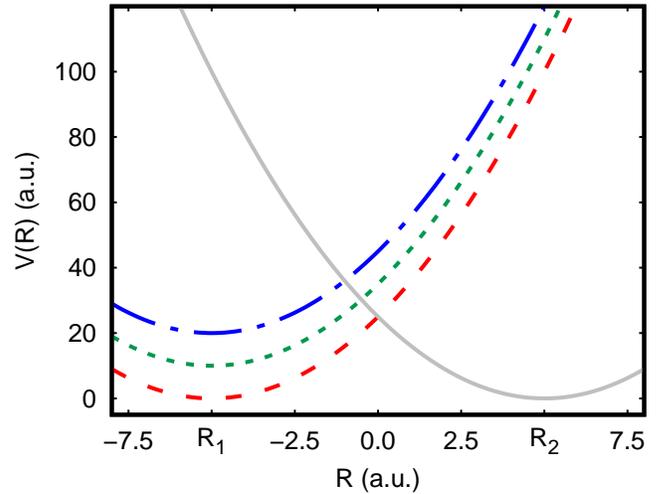}
\caption{Red (dashed) and grey (continuous) lines show 
the donor and acceptor states in model I (symmetric system); 
the green (dotted) and black (dot-dashed) lines represent the 
donor state in models II and III which are asymmetric systems 
with $\epsilon=10.0$ a.u. and $20.0$ a.u., respectively. 
$R_{1}$ and $R_{2}$ indicate the minima of the donor and acceptor states.} 
\label{fig:DApot}
\end{figure}

We find the multistate RPI for model systems with two electronic states ($\mathcal{K}=2$) 
coupled to one nuclear dof ($f=1$), and with 
three different driving forces. We note 
that all model systems are in the normal regime of 
Marcus theory where the protocol described here yields 
converged instanton solutions. 
%\sout{In the inverted regime, not considered here, 
%numerically converging to an instanton solution 
%requires imposing additional constraints.}
Diagonal elements of the diabatic potential energy matrix, $\mathbf{V}(R)$, are 
\begin{eqnarray}
    V_{ii}(R) & = & \dfrac{1}{2} M \omega^2 (R-R_{i})^2 + \epsilon \delta_{1i},
    \label{eq:diab_pot}
\end{eqnarray}
where $i=\{1,2\}$, the nuclear mass $M=2$ a.u., 
the oscillator frequency $\omega=1$ a.u. 
In Eq.~\ref{eq:diab_pot}, the Kronecker delta, $(\delta_{1i})$, 
indicates that a driving-force $\epsilon=0.0,10.0,20.0$ a.u. 
is added to the donor state (left curve), as shown in Fig.~\ref{fig:DApot}.
We choose $\beta>\beta_c$
%=1/(k_{B}T_{c})$ where $k_{B}$ is Boltzmann's constant and $T_{c}$ corresponds to the crossover
%temperature 
such that an instanton solution exists in the adiabatic limit;
we note that there is no clear analytical expression for the nonadiabatic 
crossover temperature.~\cite{Jang2001}
For the systems considered here, in the adiabatic limit ($\Delta=5$ a.u.)
we obtain a barrier frequency $\omega_{b} \approx 3$ a.u.
by fitting to an inverted parabola, and 
we use this to find
$\beta_{c}~=~2\pi/\omega_{b}~\approx~2$ a.u. for all models.~\cite{Benderskii1994} 
The electronic coupling and inverse temperature ($\beta$) values 
used here are reported in Table.~\ref{ParameterTable}.
%All parameters are provided in Table.~\ref{ParameterTable}
\begin{table}[ht!]
\renewcommand{\arraystretch}{1.25}
\begin{tabular}{|c|c|c|} \hline
 $\beta$ &\begin{tabular}[c]{@{}c@{}}Adiabatic $ \Delta $ \\ \end{tabular} & \begin{tabular}[c]{@{}c@{}} Nonadiabatic $\Delta$  \end{tabular} \\ \hline 
 4.0 & 5.0 & 6.25E-3\\ 
 3.75 & 5.33 & 6.67E-3 \\ 
 3.5 & 5.71 & 7.14E-3\\ 
 3.25 & 6.15 & 7.70E-3\\ \hline
\end{tabular}
\caption{Parameters for models I, II, and III, chosen such that $\beta\Delta$ remains constant. 
All values are in atomic units.}
\label{ParameterTable}
\renewcommand{\arraystretch}{1}
\end{table}

\section{\label{sec:level4} Simulation Details}
\subsection{\label{ssec:level4B} Optimization Algorithm}
We use the L-BFGS-B algorithm to solve the simultaneous equations for the instanton.~\cite{Zhu1997}
The L-BFGS-B algorithm is an efficient, quasi-Newton approach to optimization and 
the box constraints allow for each variable in the search to have both an upper and lower
bound. We find that a physically reasonable initial guess allows us to converge to a 
true instanton solution.

We use the following protocol to generate initial configurations. 
The minima of the two diabatic states are used 
to provide an upper and lower bound for the nuclear bead positions,
and two beads are fixed at the crossing between 
the diabatic surfaces.
We find both these constraints necessary to ensure we obtain an RPI solution
that is a first order saddle rather than a minimum. 
The electronic coordinates are constrained to lie between -1 and 1 
in all our calculations, but we find that changing this does not 
affect the outcome of the optimization.

\begin{figure}[htp]
\centering
\includegraphics[scale=0.3,angle=-90]{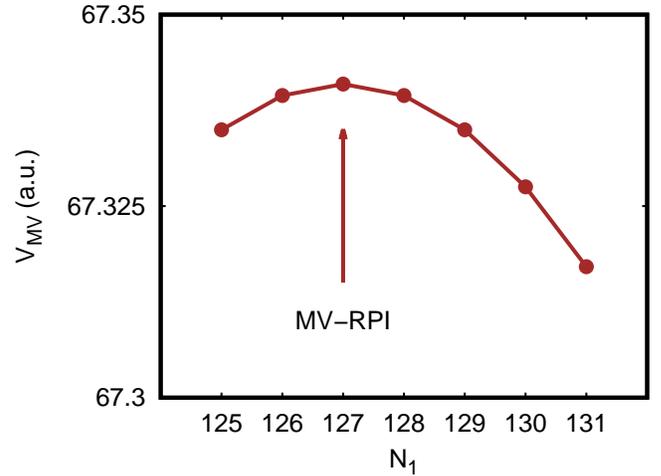} 
\caption{The effective mapping variable ring polymer potential energy 
for a 256-bead ring polymer as a function of the number of beads to the left of the crossing ($N_1$). The system shown here is symmetric 
model I with $\beta=3.25$ a.u. in the nonadiabatic regime, $\Delta =  0.0077 $ a.u.
The RPI configurations corresponds to the maximum where $N_1=N_2=127$ beads (2 beads are fixed at the crossing).}
\label{rppesN1N2}
\end{figure}

In addition to the nuclear and electronic variables, it is necessary to also optimize the ratio of number
of beads to the left and right of the crossing, $N_1/N_2$. Rather than incorporating this as an additional 
variable in our optimization algorithm, we fix this ratio for each calculation. 
The value of $N_1/N_2$ that maximizes the isomorphic ring polymer potential is 
the multistate RPI solution as shown in Fig.~\ref{rppesN1N2}.

\subsection{\label{ssec:level4A} Initial guess for the nuclear and electronic coordinates}
The optimization algorithm to find the RPI requires an initial guess. 
In theory the outcome should be independent of this guess, 
however, we find that a good initial guess is necessary to obtain a 
converged instanton solution for the large number of dofs in these calculations. 
Following previous work,~\cite{Richardson2015b} we find it necessary to fix 
two beads to the crossing point particularly for models 
with non-zero driving force where an unconstrained initial guess leads to a 
solution where all the ring polymer beads are in the lower-energy acceptor 
state rather than a first order saddle.
We initialize $N_{1}$ beads to the left (donor state) and $N_{2}$ beads 
to the right (acceptor state) of the crossing, respectively, 
such that the total number of beads of the ring polymer $N=N_1+N_2+2$.

We begin by generating initial conditions for the nuclear 
positions of the RPI according to the equations,
\begin{eqnarray}
\nonumber
&& R_j=R_{0}+\left(R_{2}-R_{0}\right) \cos\left( \dfrac{\pi i }{N_{2}+3}\right)\\
\nonumber
&&\;\;\;\;\;\;\; \text{where}\; i=1, \floor*{\dfrac{N_{2}}{2}}+1, j=1, \floor*{\dfrac{N_{2}}{2}}+1 \\
&& R_{\floor*{N_2/2}+2}=R_0 \label{IGnuc}\\
\nonumber
&& R_j=R_{0} - \left| \left( R_{0} - R_{1} \right) \cos\left( \dfrac{\pi}{2} + 
\dfrac{\pi i }{N_{1}+1}\right) \right| \\
\nonumber
&&\;\;\;\;\;\;\; \text{where} \;
i=1,N_1, j=\floor*{\dfrac{N_2}{2}}+3, 
\floor*{\dfrac{N_2}{2}}+N_{1}+2 \\
\nonumber
&& R_{\floor*{N_2/2}+N_1+3}=R_0\\
\nonumber
&& R_j=R_{0}+\left(R_{2}-R_{0}\right) \cos\left( \dfrac{\pi i }{N_{2}+3}\right) \\
\nonumber
&&\;\;\;\;\;\;\;\text{where}\;
i=\floor*{\dfrac{N_2}{2}}+4, N_2+2, j= \floor*{\dfrac{N_{2}}{2}}+N_1+4, N  
\end{eqnarray} 
Note that in the equations above, we take the lowest integer for $N_2/2$.

As a first step, for each model, we find the single 
surface RPI on the lower adiabatic surface. 
We use the nuclear positions from this RPI as the initial guess 
for our MF-RPI calculation in the adiabatic limit. 
The MF-RPI solution, in turn, provides initial nuclear configurations 
for the MV-RPI calculation, and we estimate initial values
for the electronic coordinates 
from the nuclear coordinates of the single surface instanton,
\begin{eqnarray}
\left[\mathbf{x}_{\alpha}\right]_{n} & = & \sqrt{\dfrac{e^{-\beta V_{nn}(R_{\alpha})}}{\sum_{n} e^{-\beta V_{nn}(R_{\alpha})} }} \label{eIGeqn}
\end{eqnarray}

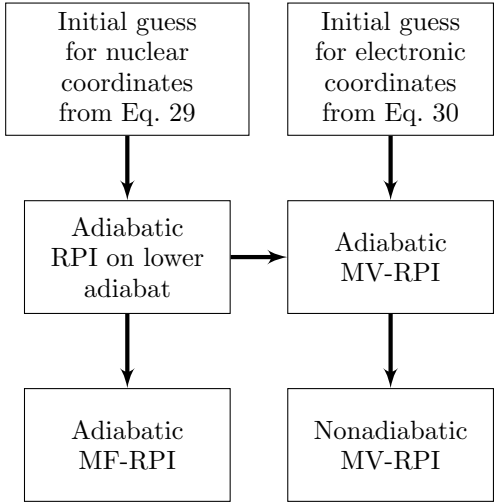
\begin{figure}[htp]
{\setstretch{1.0}\begin{tikzpicture}[auto,line/.style={draw, ultra thick, -latex', shorten >= 0pt}]
\node [rectangle, yshift=-0.5cm, xshift=-1.75cm, draw=black, fill=white, text width=3cm, minimum height=1.5cm, text centered, anchor=center] (IG) {Initial guess for nuclear \\ coordinates from Eq.~\ref{IGnuc}}; 
\node [rectangle, right of = IG, xshift=2.5cm, draw=black,  fill=white, text width=2.5cm, minimum height=1.5cm, text centered] (EEMV) {Initial guess for electronic \\ coordinates from Eq.~\ref{eIGeqn}}; 
\node [rectangle, below of = IG, yshift=-1.5cm, draw=black,  fill=white, text width=2.5cm, minimum height=1.5cm, text centered] (AASS) {Adiabatic RPI on lower adiabat}; 
\node [rectangle,  below of = AASS, yshift=-1.5cm, draw=black,  fill=white, text width=2.5cm, minimum height=1.5cm, text centered] (AAMF) {Adiabatic MF-RPI}; 
\node [rectangle, right of = AASS, xshift=2.5cm, draw=black,  fill=white, text width=2.5cm, minimum height=1.5cm, text centered] (AAMV) {Adiabatic MV-RPI};
\node [rectangle, below of = AAMV, yshift=-1.5cm, draw=black, fill=white, text width=2.5cm, minimum height=1.5cm, text centered] (NAMV) {Nonadiabatic MV-RPI}; 
\begin{scope}[every path/.style=line]
  \path (IG) -- (AASS);
  \path (AASS) -- (AAMF);
  \path (AASS) -- (AAMV);
  \path (EEMV) -- (AAMV);
  \path (AAMV) -- (NAMV);
\end{scope}
\end{tikzpicture}
\caption{We start with an initial nuclear configuration generated according to Eq.~\ref{IGnuc} 
 and find the single surface RPI on the lower adiabatic state. The initial electronic 
 configurations are generated according to Eq.~\ref{eIGeqn}. 
 The arrows connect initial guess configurations with optimized RPI configurations.
} \label{fig:Protocol}}
\end{figure}
The flowchart in Fig.~\ref{fig:Protocol} outlines the procedure
used to generate initial configurations where we use the fact that 
the single surface RPI, the MF-RPI, and the MV-RPI should have 
identical nuclear configurations in the adiabatic limit. 
The nonadiabatic RPI calculations employ the adiabatic RPIs 
as the initial guess for nuclear configurations in all cases.

\subsection{\label{ssec:level4C}
Nonadiabatic and Adiabatic Reaction Rate Theories}
For a two-level system where the reactant and product 
potential surfaces are displaced harmonic oscillators,
the nonadiabatic reaction rate according to 
Fermi's Golden Rule (FGR) rate is
\begin{eqnarray}
k_{\mathrm{FGR}} & = & \dfrac{2\pi}{\omega} |\Delta|^2 
e^{\nu \mathrm{z} - S\mathrm{coth(z)}} I_{\nu}(S\mathrm{csch(z)}),
\label{kFGR}
\end{eqnarray}
where $\mathrm{z} = \beta\omega/2, \nu = \epsilon/\omega, 
S = M\omega(R_{r}-R_{p})^2/2$, and $I_{\nu}$ is a modified 
Bessel function of the first kind. $R_{r}$ and $R_{p}$ are 
minima of the reactant and product oscillators, and $\omega$ 
is their frequency. 

In the adiabatic limit, the single surface RPI 
rate~\cite{Richardson2009} on the lower adiabat is:
\begin{eqnarray}
k_{\mathrm{RP}} & \approx & \dfrac{e^{-\beta_{N}
V_{\mathrm{RP}}(\tilde{\mathbf{R}}_{\alpha})} }
{\mathcal{Z}_{r} \beta_{N}} 
\left(\dfrac{M z_{N}}{2\pi\beta_{N}}\right)^{\frac{1}{2}} 
\sideset{}{'}
\prod_{\alpha=1}^{N}\dfrac{1}{\beta_{N}|\lambda_{\alpha}|} 
\label{kRP}
\end{eqnarray}
where 
$z_{N} = \sum_{\alpha}(\tilde{\mathbf{R}}_{\alpha}-\tilde{\mathbf{R}}_{\alpha+1})^2$,
and as in Sec.~\ref{ssec:level2E}, the $^\prime$ indicates $\alpha=2$ 
is excluded from the product.

\section{\label{sec:level5}Results}
\begin{figure}[htp]
\centering
\includegraphics[scale=0.3,angle=-90]{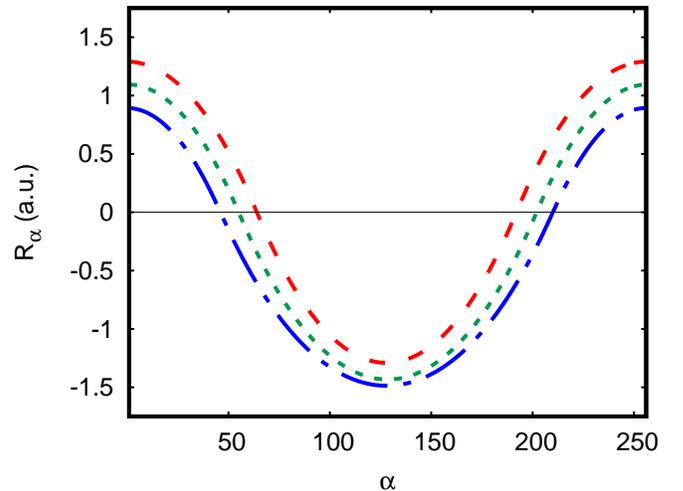}
\caption{MV-RPI nuclear bead positions as a function of  bead number 
for model I (red dashed line), model II (green dotted line), and model III (blue dot-dashed line) 
at $\beta=3.25$ a.u. with $\Delta=0.0077$ a.u. (nonadiabatic).
The horizontal black line indicates the position at which donor and acceptor states cross.
Note that bead positions for models II and III have all been shifted by $0.5$ a.u. and $1.0$ a.u., respectively, 
so that the crossing of all the three models coincide. The number of beads to the left of the crossing increases
as the driving force increases.} \label{fig:AdiaInsAllModels}
\end{figure}

\begin{figure}[htp]
\centering
\includegraphics[scale=0.3,angle=-90]{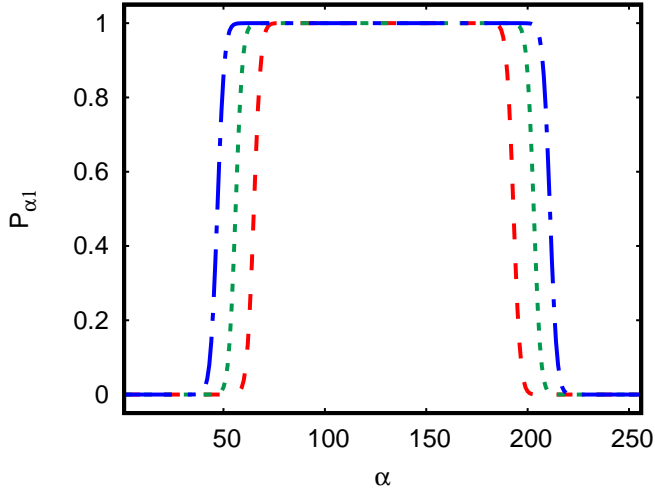}
\caption{MV-RPI bead populations as a function of  bead number in the donor state for 
model I (red dashed line), model II (green dotted line), and model III (blue dot-dashed line) 
at $\beta=3.25$ a.u. with $\Delta=0.077$ a.u. (nonadiabatic). 
In keeping with the nuclear position 
changes along the instanton path, the number of beads in the donor state increases 
as the driving force increases.}
\label{fig:DSPopAdiaMV}
\end{figure}

We present the MV-RPI for models I, II, and III converged with $256$ beads at 
$\beta=3.25$ a.u for a nonadiabatic case with $\Delta=0.0077$ a.u..
We find that as the driving force increases the number of beads to the left of the crossing 
also increases as shown by the nuclear MV-RPI in Fig.~\ref{fig:AdiaInsAllModels}.
For the electronic MV-RPI, we use the normalized Wigner population estimator to 
calculate the population of the $\alpha^{\mathrm{th}}$ bead in the 
$n^{\mathrm{th}}$ state,
\begin{eqnarray}
    \mathcal{P}_{\alpha n} = \left[\mathbf{x}_{\alpha}\right]_{n}^2.
    \label{popWig}
\end{eqnarray}
We show that the electronic MV-RPI in Fig.~\ref{fig:DSPopAdiaMV}, in agreement with
the nuclear RPI, show an increase in the number of beads in the donor
state as the driving force increases.
Table.~\ref{table:MV} summarizes the key results of the effect of the driving force on the MV-RPI:
the ratio $N_{1}/N_{2}$ grows as a function of the driving force, and is 
independent of the coupling strength.
\begin{table}[h]
\renewcommand{\arraystretch}{1.5}
\centering
\begin{tabular}{|c|c|c|c|c|c|c|c|c|c|} \hline
\multicolumn{4}{|c|}{} & \multicolumn{3}{c|}{Adiabatic}     & \multicolumn{3}{c|}{Nonadiabatic}  \\ \hline
$\epsilon$ & $N_{1}$ & $N_{2}$& $R_{0}$ & $\bar{R}$ & $\mathbb{P}_{1}$ & $\mathbb{P}_{2}$ & $\bar{R}$ & $\mathbb{P}_{1}$ & $\mathbb{P}_{2}$ \\ \hline
0.0 & 127  & 127 &  0.0 & $10^{-5}$  &  0.50&  0.50 & $10^{-5}$ &   0.50 &  0.50 \\ \hline
10.0 & 146  & 108 & -0.5 & -0.71  & 0.57 & 0.43 & -0.74 &  0.57 &  0.43 \\ \hline
20.0 & 163 & 91 &  -1.0 &  -1.34 & 0.63 & 0.37 &  -1.45  &  0.65   & 0.35  \\ \hline
\end{tabular}
\caption{256-bead MV-RPI for models I, II and III with $\beta=3.25$ a.u. for 
$\Delta=6.15$ a.u. (adiabatic) and $\Delta=0.0077$ a.u. (nonadiabatic). 
We report the number of beads on the 
donor ($N_1$) and acceptor ($N_2$) surfaces excluding the two beads 
constrained to the crossing. In each case, we also tabulate the 
position of the nuclear centroid $(\bar{R})$ and bead-averaged 
donor and acceptor state populations, $\mathbb{P}_1$ and $\mathbb{P}_2$, 
respectively. 
The values of the nuclear variables, and the energies are given in atomic units.} \label{table:MV}
\renewcommand{\arraystretch}{1.0}
\end{table}

\begin{figure}[htp]
\centering
\includegraphics[scale=0.3, angle=-90]{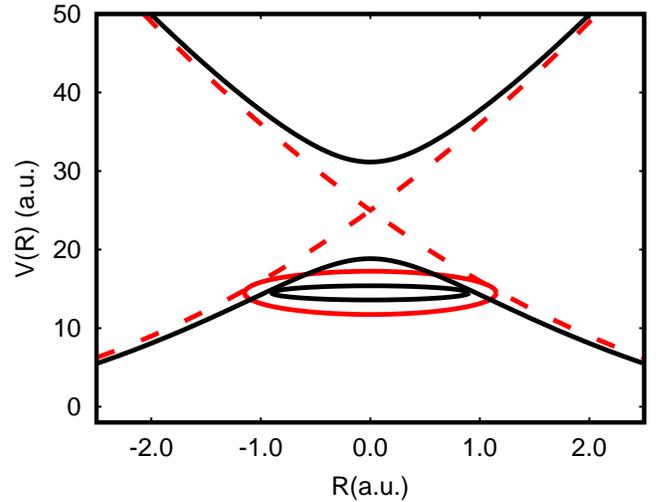}
\caption{The adiabatic potential energy surfaces in the adiabatic (black line) 
and nonadiabatic (red dashed line) coupling limits for model I at $\beta = 3.25$ a.u.
are shown here. A cartoon repsentation of the extent of spread in the nuclear instanton for both the nonadiabatic (red outer ellipse) and adiabatic (black inner ellipse) limits is also shown.} \label{fig:NAAPot}
\end{figure}

\begin{figure}[htp]
\centering
\includegraphics[scale=0.3, angle=-90]{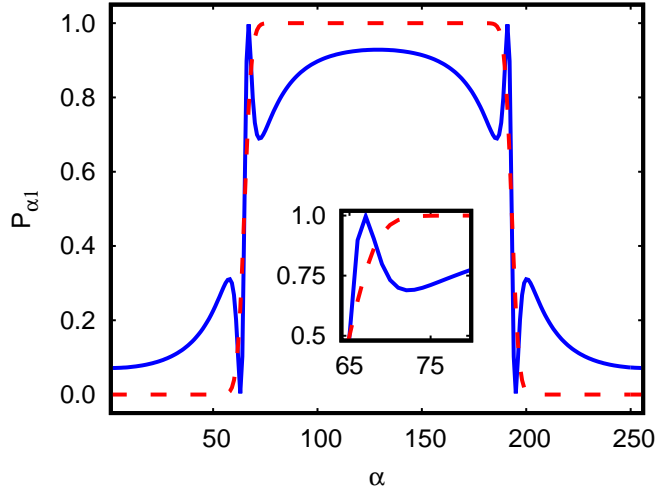}
\caption{electronic MV-RPI donor state populations as a function of  bead number for model I with $\beta=3.25$ a.u.  and $\Delta=6.15$ a.u. (adiabatic, blue line),
and $\Delta=0.0077$ a.u. (nonadiabatic, red dashed line). The inset illustrates 
electronic state populations vary gradually in the nonadiabatic case
but exhibit some oscillatory structure in the adiabatic regime.} \label{fig:NAAPop} 
\end{figure}

Next, we explore the structure of the MV-RPI for model I 
in the adiabatic ($\Delta=6.15$ a.u.) and 
nonadiabatic ($\Delta=0.0077$ a.u.) regimes with $\beta=3.25$ a.u.
\begin{figure}[htp]
\centering
\includegraphics[scale=0.3, angle=-90]{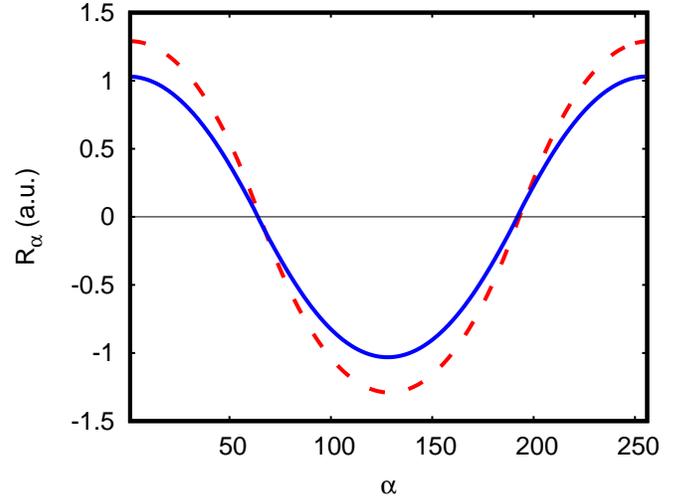}
\caption{
    MV-RPI nuclear bead positions as a function of  bead number for 
    model I with $\beta=3.25$ a.u. and $\Delta=6.15$ a.u. (adiabatic,
    blue line), and $\Delta=0.0077$ a.u. (nonadiabatic, red dashed line) . 
    The nonadiabatic instanton exhibits a wider spread than the adiabatic 
    instanton in keeeping with lower curvature of the barrier in the latter
    case. The black horizontal line marks the position at which donor 
    and acceptor state diabats cross.} \label{fig:NAAIns}
\end{figure}
In keeping with findings from other multistate instanton calculations,~\cite{Cao1995} we
find that the spread of the nonadiabatic nuclear MV-RPI, 
shown in Fig.~\ref{fig:NAAIns}, is wider than 
the corresponding adiabatic MV-RPI. This is a consequence
of the higher curvature of the barrier in the nonadiabatic
limit as shown in Fig.~\ref{fig:NAAPot}.
The electronic MV-RPI (donor state populations) are shown in Fig.~\ref{fig:NAAPop}. 
The two beads at the crossing have equal population in the donor and acceptor electronic 
states in both the adiabatic and nonadiabatic models. However, as shown in the inset,
the state populations exhibit oscillatory features near the crossing in the adiabatic limit, 
a feature arising from our use of a diabatic state representation. 
We also find that while the golden-rule instanton places individual 
beads on one or the other surface,~\cite{Richardson2015b} the MV-RPI 
allows for beads with partial populations in keeping with the quantum 
mechanical probability of the system being in a particular electronic state
along the instanton path.

\begin{figure}[htp]
\centering
\includegraphics[scale=0.3,angle=-90]{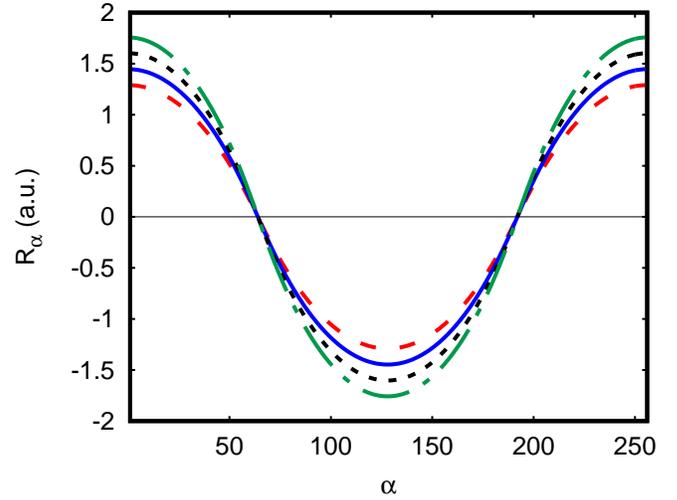}
\caption{Nucelar bead positions as a function of bead number for a 256-bead 
MV-RPI for model I with $\Delta=0.00625$ a.u. (nonadiabatic) and $\beta = 3.25, 
3.5, 3.75, 4.0$ a.u. represented by red dashed, blue continuous, black dotted, and green dot-dashed lines, respectively. The horizontal line is the crossing of the donor and acceptor diabats.}  \label{fig:betaVaryIns}
\end{figure}

Fig.~\ref{fig:betaVaryIns} shows the effect of temperature on the nuclear 
MV-RPI.  Specifically, for model I with $\Delta=0.00625$ a.u., we vary 
$\beta$ from $3.25$ a.u. to $4.0$ a.u. in steps of $0.25$ a.u. 
We find the nuclear MV-RPI spread increases as temperature decreases;
as the system is cooled, we expect a shift towards deep tunneling 
with a corresponding increase in the extent of the instanton. 
The electronic MV-RPI does not change significantly over the range
of temperatures considered here. 

\begin{figure}[htp]
\centering
\includegraphics[scale=0.3, angle=-90]{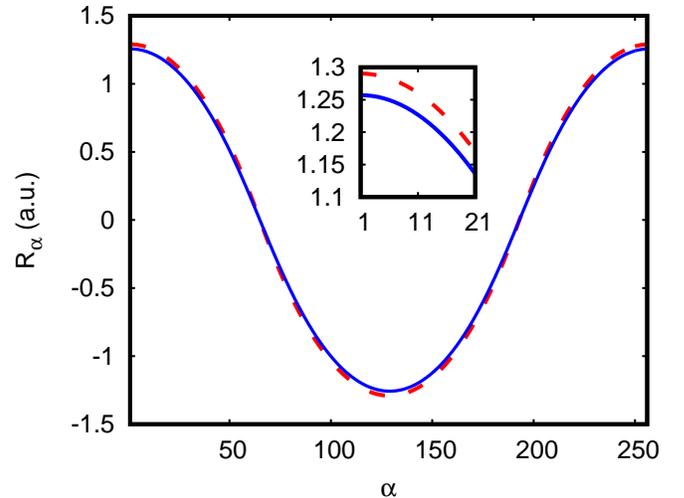}
\caption{Nuclear bead positions as a function of bead number for a 256-bead MV-RPI determined for model I at $\beta = 3.25$ a.u. in the nonadiabatic limit ($\Delta = 0.0077$ a.u.).
The MF-RPI nuclear positions are represented using the blue line while those obtained from the MV-RPI 
using the red dotted lines. The inset highlights nuclear positions for the first 20 beads where the 
small numerical difference between the two instantons is most noticeable.}\label{fig:MFcmvRPI}
\end{figure}

Next, we compare the MF-RPI and the MV-RPI. 
Given that both are derived from equivalent,
exact representations of the quantum canonical 
partition function, we expect the two nuclear instantons
to be nearly indistinguishable as shown in Fig.~\ref{fig:MFcmvRPI}.
%the nuclear instantons are nearly identical in the nonadiabatic case and 
%identical in the adiabatic case (not pictured). 

\begin{figure}[htp]
\centering
\includegraphics[scale=0.27,angle=-90]{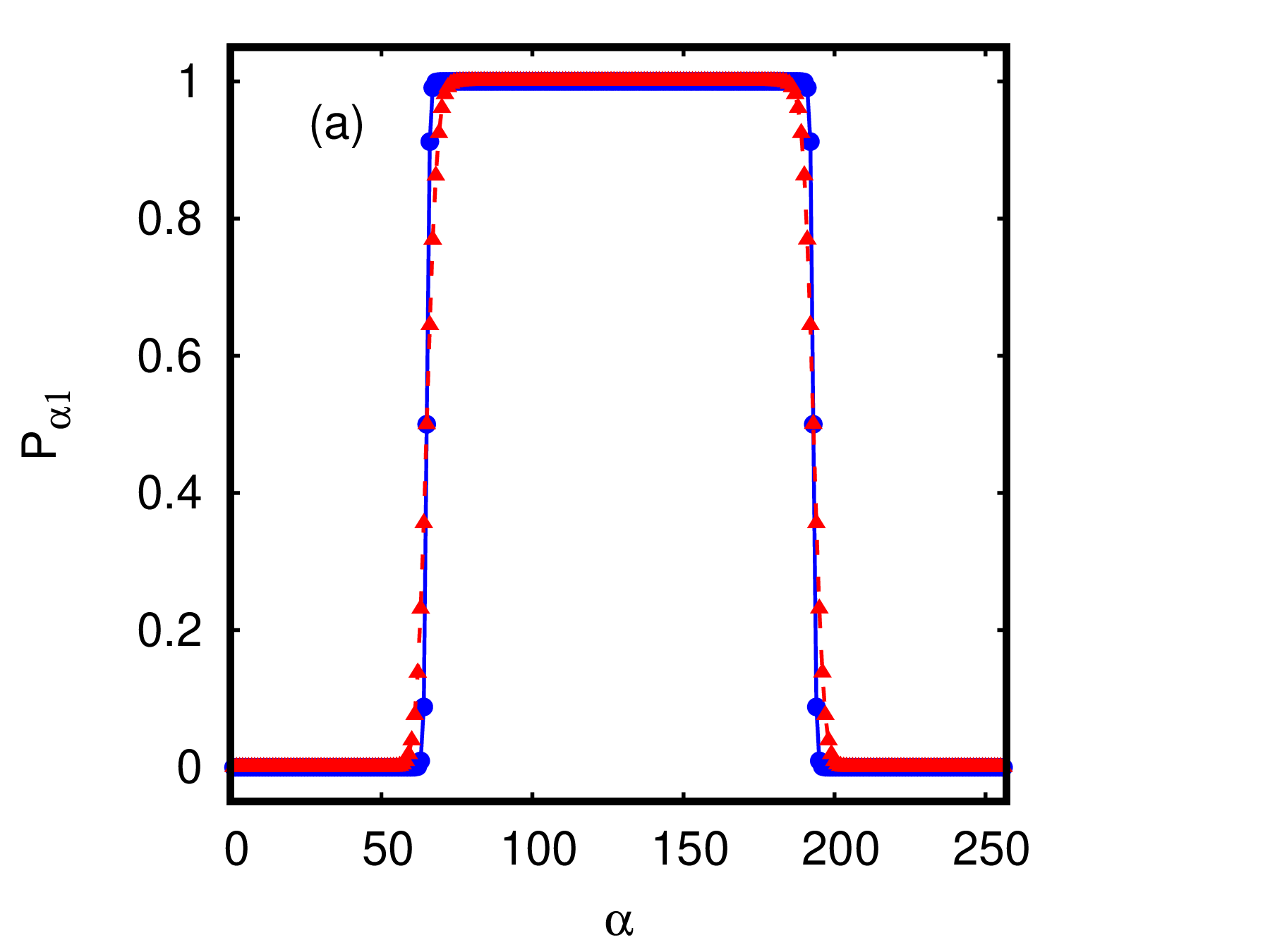}\\
\includegraphics[scale=0.27,angle=-90]{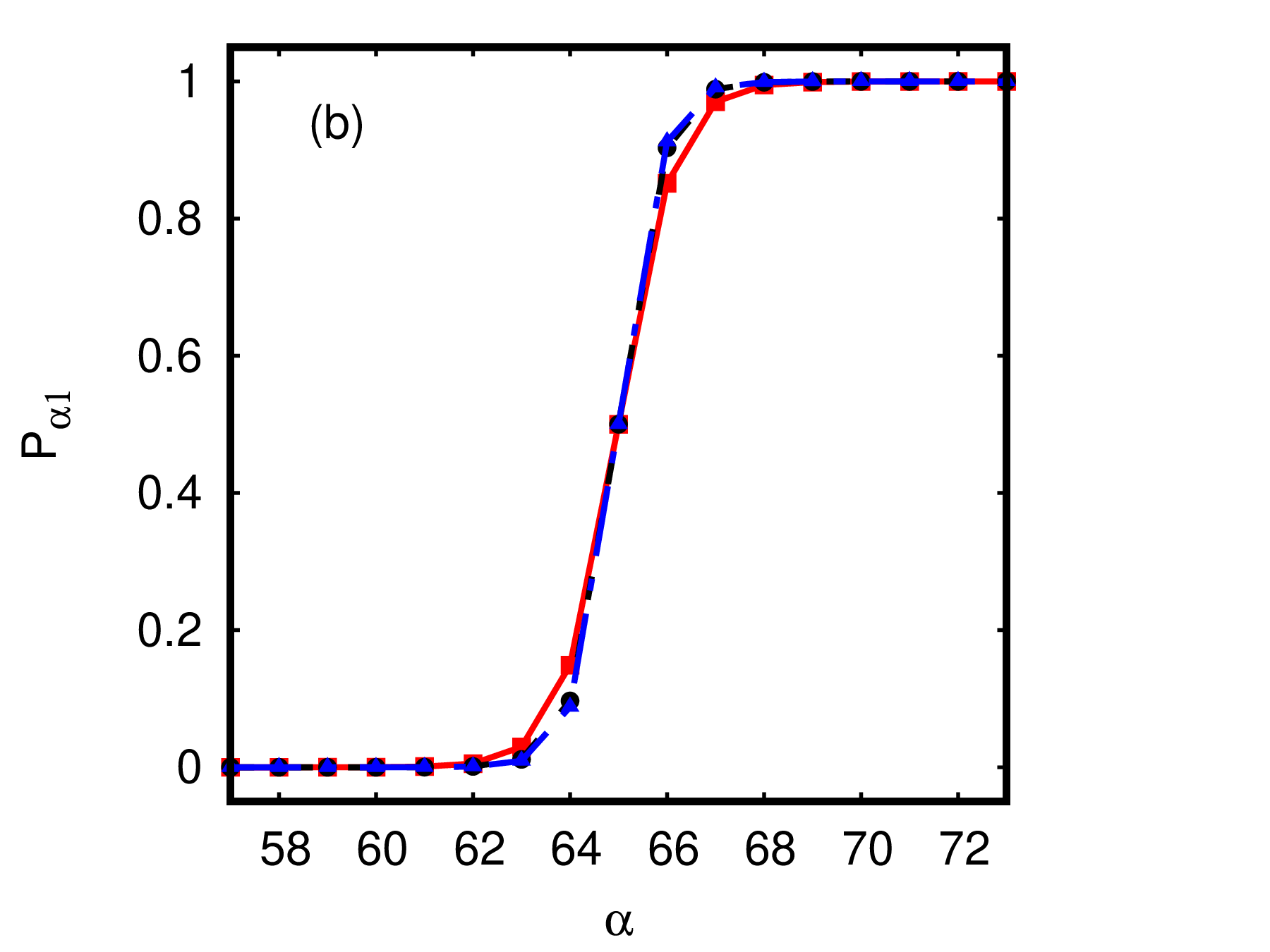}\\
\includegraphics[scale=0.27,angle=-90]{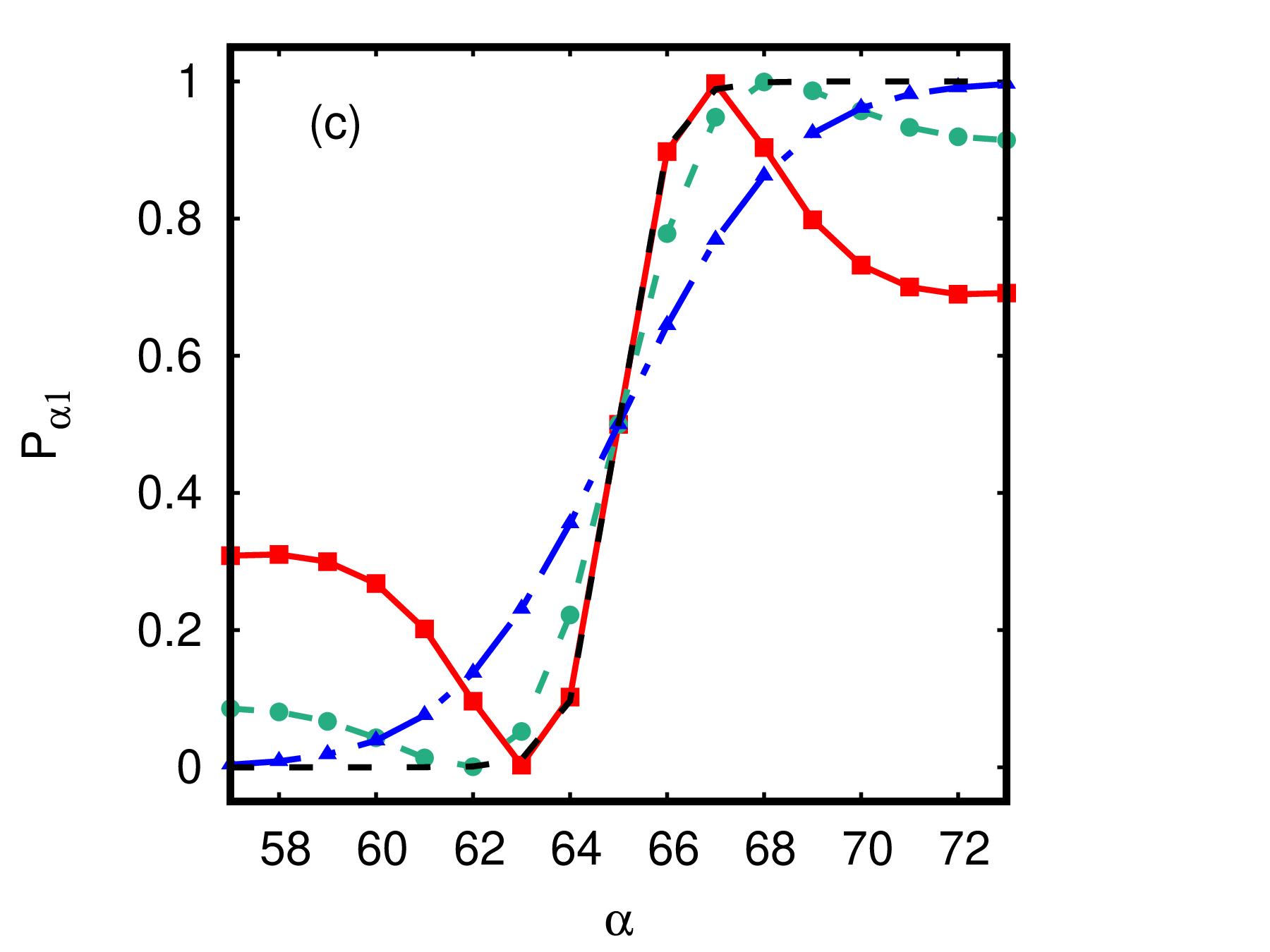}
\caption{
(a) Comparing donor state populations obtained from the nuclear MF-RPI instanton (blue line with circles) 
against the electronic MV-RPI populations (red dashed line with triangles) for model I with $\Delta=0.001$ a.u. and $\beta=3.25$ a.u.
(b) Populations for beads in the vicinity of the crossing obtained 
from the nuclear MF-RPI instanton for model I 
with $\beta=3.25$ a.u. shown for three
different coupling strengths, $\Delta=6.15$ a.u (red line with squares), $\Delta=2.5$ a.u (black dashed line with triangles),
and $\Delta=0.001$ a.u (blue dot-dashed line with circles). 
(c) Electronic MV-RPI populations for beads in the vicinity of the crossing for model I 
with $\beta=3.25$ a.u. for three 
different coupling strengths, $\Delta=6.15$ a.u (red line with squares), $\Delta=2.5$ a.u (green dashed line with circles),
and $\Delta=0.001$ a.u (blue dot-dashed line with triangles). 
For comparison, the MF-RPI populations are shown for $\Delta=2.5$ a.u. (black dashed line).}

\label{fig:MFmvPop}
\end{figure}

Although the MF-RPI does not include explicit electronic variables, it is possible
to estimate the electronic state populations from the nuclear bead positions along
the nuclear instanton path. In keeping with previous work~\cite{Schwieters1998},
we calculate the donor state population of each bead using the following {\it ad hoc}
expression,
\begin{eqnarray}
\mathbb{P}_{\alpha 1} = \dfrac{e^{-\beta V_{11}(\mathbf{R}_{\alpha})}}{e^{-\beta V_{11}(\mathbf{R}_{\alpha})} + e^{-\beta V_{22}(\mathbf{R}_{\alpha})}} \label{MFpop}
\end{eqnarray}
We compare this MF-RPI result against the electronic MV-RPI populations in 
Fig.~\ref{fig:MFmvPop}(a) for nonadiabatic model I with $\Delta=0.001$ a.u 
and $\beta=3.25$ a.u..
Examining beads in the vicinity of 
the crossing, shown in Fig.~\ref{fig:MFmvPop}(b) 
and Fig.~\ref{fig:MFmvPop}(c), we find that populations obtained 
from the nuclear MF-RPI fail to distinguish between the adiabatic
and nonadiabatic regimes over three orders of magnitude. 
However, the electronic MV-RPI populations show 
a smooth transition from beads on one state to the other along 
the instanton path in the nonadiabatic case and an oscillatory structure
in the adiabatic case (as expected when working in the diabatic representation).
We also find that in the adiabatic regime, beads away from the crossing 
are not fully in one or the other diabatic state in keeping with the
underlying model.

\begin{figure}[htp]
\includegraphics[scale=0.3, angle=-90]{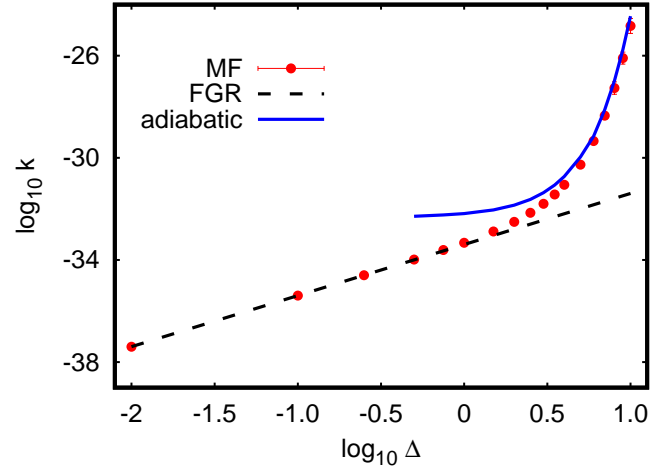}
\caption{MF-RPI rate constants (red circles) 
as a function of the coupling strength for model I 
at $\beta=4.0$ a.u.
In the nonadiabatic regime, we find good agreement
with the FGR rate constants (dashed black line) 
and in the adiabatic regime, MF-RPI results are 
within a factor of 2 of the single surface 
RPI rate constants (blue line).
} \label{fig:rates}
\end{figure}

Finally, we calculate MF-RPI rate constants. 
In Fig.~\ref{fig:rates}, we present results 
for model I as a function of the electronic 
coupling, $\Delta$. In the nonadiabatic regime, we find the 
MF-RPI rate constants are in good agreement with FGR rate
constants. As we approach the adiabatic, strong coupling
regime where FGR is not applicable, we calculate the 
single surface RPI rate on the lower adiabatic surface
and find that they agree well with the MF-RPI rates, 
as reported in Table~\ref{table:ratesCoupling}.
In the intermediate coupling regime, we 
find the MF-RPI rate expression interpolates smoothly
and accurately 
between the nonadiabatic and adiabatic dynamic regimes.

\begin{table}[htp]
\renewcommand{\arraystretch}{1.4}
\begin{tabular}{|c|c|c|c|} \hline
$\Delta$ & $\log_{10} k_{\mathrm{MF-RPI}}$ & $\log_{10} k_{\mathrm{FGR}} $ & $\log_{10}k_{RP}$ \\ \hline
0.01 & -37.398 & -37.395 & - \\ 
0.1 & -35.397  & -35.395 &  - \\ 
0.25 &  -34.598 & -34.599  &  - \\
1.0 & -33.327 & -33.395  &  -32.182 \\
2.5 &  -32.153 & -32.599  & -31.628 \\
5.0 & -30.269 & -31.997  &  -29.966 \\
10.0 & -24.836 & -31.395 & -24.444 \\ \hline
\end{tabular}
\caption{Comparing rate constants for model I  at $\beta=4.0$ a.u. with 
different coupling strengths. $k_{\mathrm{MF-RPI}}$ is the MF-RPI 
rate constant, $k_{\mathrm{FGR}}$ is the rate constant obtained 
using Fermi's Golden Rule, and $k_{\mathrm{RP}}$ is the 
rate constant computed using single surface RPI determined 
on the lower adiabatic surface. Note that $k_{\mathrm{RP}}$ 
is only calculated for cases where the RPI has a zero mode.} \label{table:ratesCoupling}
\renewcommand{\arraystretch}{1}
\end{table}

We also find good agreement between MF-RPI rates
and FGR rates for models I, II, and III in 
nonadiabatic regime with $\Delta=0.0625$ a.u. 
where driving force is varied 
as reported in Table~\ref{table:ratesDF}.

\begin{table}[htp]
\renewcommand{\arraystretch}{1.4}
\begin{tabular}{|c|c|c|c|} \hline
Model & $\epsilon$ & $\log_{10} k_{\mathrm{MF-RPI}} $ & $\log_{10} k_{\mathrm{FGR}}$ \\ \hline
I & 0.0 & -37.805 & -37.803 \\ 
II & 10.0 &  -29.913 & -29.913  \\ 
III & 20.0 & -23.485 &  -23.508 \\ \hline
\end{tabular}
\caption{MF-RPI rates compared with FGR rates for 
nonadiabatic models I, II and III with 
$\beta=4.0$ a.u., $\Delta=0.0625$ a.u. } \label{table:ratesDF}
\renewcommand{\arraystretch}{1}
\end{table}

\section{\label{sec:level6}Conclusion}
In this paper, we compute multistate RPIs for nonadiabatic systems
that are valid in the adiabatic regime as well. We show that, as expected, 
both the MF and MV representations arrive at identical 
nuclear instantons, but the electronic MV-RPI uniquely reports on electronic
state transitions along the instanton path. We obtain analytic expressions 
for the zero-mode in both cases and numerically establish that all solutions 
reported here correspond to a first order saddle.
%We characterize changes in the multistate instantons as driving force increases 
%and as temperature decreases.
%We find that the nuclear multistate RPIs show an increase in the number of beads
%in the donor state as driving force increases. 
We find that the electronic MV-RPI shows
a gradual change in population of beads from donor to acceptor state
in the nonadiabatic case and a more steep change in the adiabatic case
along with some oscillatory structure. We clearly demonstrate
that electronic state populations obtained from the nuclear MF-RPI 
fail to distinguish between the adiabatic and nonadiabatic regimes.
We also note that the electronic MV-RPI captures beads 
with fractional state population in keeping with the quantum mechanical 
probability of being in a given electronic state, unlike instantons based 
on Fermi's golden rule that place beads on one or the other surface.

Finally, we obtain an expression for the MF-RPI rate and demonstrate
that it is accurate for the calculation of rates in both the adiabatic
and nonadiabatic regimes. We expect this will prove a powerful tool 
for rapid rate calculations. 
We do not provide a separate MV-RPI rate equation since the MF
and MV path integral expressions are equivalent (and exact) for
the calculation of statistical properties, and as such can be 
reasonably expected to yield 
numerically equivalent instanton rates in all regimes.
Rather, we view the MV-RPI as an essential step towards characterizing 
the mechanism of nonadiabatic dynamic processes through its connection 
with an optimal dividing surface transition state theory~\cite{Richardson2009} 
and as a way to initialize trajectories in MV-RPMD simulations of multistate 
population dynamics.

\begin{acknowledgments}
The authors would like to thank Prof. Ezra and Prof. DiStasio for helpful discussions. This work was funded through a National 
Science Foundation CAREER grant (Award No. CHE-1555205).
\end{acknowledgments}

\begin{appendices}
\section{\label{sec:AppendA} Zero mode of the MF instanton}
We obtain the MF action from the continuous ($N\rightarrow\infty$) limit of 
the effective MF ring polymer potential
\begin{eqnarray}
\mathcal{S}_{\mathrm{MF}} = && \lim_{N\rightarrow\infty}\beta_{N}V_{\mathrm{MF}} \nonumber \\
= && \lim_{N\rightarrow\infty} \beta_{N}\left[ \dfrac{M}{2\beta_{N}^2}\sum_{\alpha} (\mathbf{R}_{\alpha} - \mathbf{R}_{\alpha+1})^2 - \dfrac{1}{\beta_{N}} \ln \Gamma_{\mathrm{MF}} \right]\nonumber \\
 = && \int_{0}^{\beta} d\tau \left[ \dfrac{M}{2}\left( \dfrac{d \mathbf{R}(\tau)}{d\tau}\right)^2 - \dfrac{d \ln| \Gamma_{\mathrm{MF}}[\mathbf{R}(\tau)]| }{d\tau}  \right] \nonumber \\
 = && \int_{0}^{\beta} d\tau \left[ \dfrac{M}{2}\left( \dfrac{d \mathbf{R}(\tau)}{d\tau}\right)^2 + \mathcal{V}_\mathrm{{MF}}\right] 
 \label{eq:mf_action}
\end{eqnarray}
Here, we define $ -\dfrac{d \ln |\Gamma_{\mathrm{MF}}[\mathbf{R}(\tau)] |}{d\tau} \equiv \mathcal{V}_\mathrm{{MF}}$ for clarity of  presentation.
Note that in obtaining Eq.~\ref{eq:mf_action}, we used the finite difference definition
of a derivative for the nuclear term and we introduce the integral of a differential
operator in $\tau$ for the $\ln |\Gamma_\mathrm{MF}|$ term.
Setting the first variation of the action to zero, 
we obtain Newton's equations in imaginary
time on the inverted potential,
\begin{eqnarray}
\delta \mathcal{S}_{\mathrm{MF}} =
- M \dfrac{d^2 \mathbf{R}(\tau)}{d\tau^2} + \nabla_{\mathbf{R}} \mathcal{V}_\mathrm{{MF}} = 0.
\label{MFeom} 
\end{eqnarray}
The MF instanton is the solution to these $N\times f$ equations.

The instanton is an unstable periodic orbit, and this is established
by calculating the eigenvalues of the stability matrix. 
We obtain the stability matrix from the second 
variation of the action $\delta^2 \mathcal{S}_{\mathrm{MF}}$,
\begin{eqnarray}
\Lambda_{\mathrm{MF}} \equiv - M\dfrac{d^2}{d\tau^2} + \nabla_{\mathbf{R}}\nabla_{\mathbf{R}}^{T}\mathcal{V}_\mathrm{{MF}}.
\end{eqnarray}
Differentiating Eq.~\ref{MFeom} with respect to imaginary time, we obtain
\begin{eqnarray}
&& \dfrac{d}{d\tau} \left[ - M \dfrac{d^2 \mathbf{R}(\tau)}{d\tau^2} + \nabla_{\mathbf{R}}\mathcal{V}_\mathrm{{MF}}  \right] = 0  \nonumber \\
&& \left[ - M \dfrac{d^2 }{d\tau^2} + \nabla_{\mathbf{R}}\nabla_{\mathbf{R}}^{T}\mathcal{V}_\mathrm{{MF}} \right] \dot{\mathbf{R}}(\tau) = 0 \times \dot{\mathbf{R}}(\tau).
\label{eq:stab_mat_evec}
\end{eqnarray}
It is clear that Eq.~\ref{eq:stab_mat_evec} is the eigenvalue equation 
corresponding to operator $\Lambda_{\mathrm{MF}}$ and the zero-mode 
is the velocity mode, $\dot{\mathbf{R}}(\tau)$, with a zero eigenvalue. 

\section{\label{sec:AppendB} Zero mode of the MV instanton}
The action $\mathcal{S}_{\mathrm{MV}}$ is the imaginary-time 
integral of the Langrangian $\mathcal{L}_{\mathrm{MV}}$ given by
\begin{eqnarray} 
\mathcal{L}_{\mathrm{MV}} & = & \left[ \makecell{ \dfrac{M}{2} \left( \dfrac{d \mathbf{R}(\tau)}{d\tau} \right)^{T}\left( \dfrac{d \mathbf{R}(\tau)}{d\tau} \right) + \dfrac{d}{d\tau} \mathbf{x}(\tau)^{T}\mathbf{x}(\tau) \vspace*{0.25cm} \\ - \dfrac{d}{d\tau} \ln \left| \Gamma[\mathbf{R}(\tau),\mathbf{x}(\tau)] \right| } \right]  \nonumber \\
\label{eq:mv_lagrangian}
\end{eqnarray}

Following our treatment of the MF action, we define 
\begin{equation}
\mathcal{V}_{\mathrm{MV}} \equiv - \dfrac{d}{d\tau} \ln \left| \Gamma[\mathbf{R}(\tau),\mathbf{x}(\tau)] \right| \notag
\end{equation}
The first variation of action yields a set of coupled equations of motion for the nuclear and electronic variables,
\begin{eqnarray}
\delta \mathcal{S}_{\mathrm{MV}} & = & \int d\tau\ \delta\mathbf{R}(\tau) \left[ - M \left( \dfrac{d^2 \mathbf{R}(\tau)}{d\tau^2} \right) + \nabla_{\mathbf{R}} \mathcal{V}_{\mathrm{MV}} \right]  \notag \\
& + & \int d\tau\ \delta \mathbf{x}(\tau) \left[  - \nabla_{\mathbf{x}} \mathcal{V}_{\mathrm{MV}} \right].
\label{MVeom}
\end{eqnarray}
The second variation of action gives,
\begin{eqnarray}
\delta^2 \mathcal{S}_{\mathrm{MV}} & = & \int d\tau \left[ \begin{matrix}
\delta \mathbf{R}(\tau) & \delta \mathbf{x}(\tau) \end{matrix} \right] \Lambda_\mathrm{MV} \left[\begin{matrix} \delta \mathbf{R}(\tau) \vspace*{0.25cm} \\ \delta \mathbf{x}(\tau) \end{matrix} \right], \nonumber
\end{eqnarray}
where $\Lambda_{\mathrm{MV}}$ is the $(f+\mathcal{K})\times (f+\mathcal{K})$ 
dimensional stability matrix,
\begin{eqnarray}
\left[\begin{matrix} - M \left( \dfrac{d^2 }{d\tau^2} \right) + \nabla_{\mathbf{R}} \nabla_{\mathbf{R}}^{T} \mathcal{V}_{\mathrm{MV}} &  - \nabla_{\mathbf{R}} \nabla_{\mathbf{x}}^{T} \mathcal{V}_{\mathrm{MV}}\vspace*{0.25cm} \\ 
- \nabla_{\mathbf{x}} \nabla_{\mathbf{R}}^{T} \mathcal{V}_{\mathrm{MV}} & - \nabla_{\mathbf{x}} \nabla_{\mathbf{x}}^{T} \mathcal{V}_{\mathrm{MV}}\end{matrix} \right].  \nonumber \\
\end{eqnarray}

Differentiating  Eq.~\ref{MVeom}, we once again obtain an eigenvalue 
equation for the stability operator,
\begin{eqnarray}
\Lambda_{\mathrm{MV}} \left[\begin{matrix} \dot{\mathbf{R}}(\tau) \vspace*{0.25cm} \\ \dot{\mathbf{x}}(\tau)  \end{matrix} \right] = \left[\begin{matrix} 0 \vspace*{0.25cm} \\ 0 \end{matrix} \right] = 0 \left[\begin{matrix} \dot{\mathbf{R}}(\tau) \vspace*{0.25cm} \\ \dot{\mathbf{x}}(\tau) \end{matrix} \right].
	\label{eq:zeromode_mv}
\end{eqnarray}

It is clear from Eq.~\ref{eq:zeromode_mv} 
that the collective velocity mode (in both nuclear and 
electronic variables) is the zero mode. 

\end{appendices} 
\section*{References}

\bibliography{Non-adiabatic-Instanton}

\end{document}